\documentstyle[12pt,a4]{article}

\newcommand{\x}{{\rm x}}

\newcommand{\dr}{\stackrel{\leftarrow}{\delta}}
\newcommand{\dl}{\stackrel{\rightarrow}{\delta}}

\newcommand{\df}{\delta Y^A({x})}
\newcommand{\da}{\delta Y_A^*({x})}
\newcommand{\s}{{\cal S}}
\newcommand{\Se}{S_{ext}[Y,Y^*]}
\newcommand{\Sep}{S_{ext}^{\psi}[Y,Y^*]}
\newcommand{\Sext}{S_{ext}[\eta,a,c;\eta^*,a^*,c^*]}
\newcommand{\Sextt}{S_{ext}[\eta,a,c,\bar{c},b;\eta^*,a^*,c^*,\bar{c}^*,b^*]}
\newcommand{\Sexttp}
{S_{ext}^{\psi}[\eta,a,c,\bar{c},b;\eta^*,a^*,c^*,\bar{c}^*,b^*]}

\newcommand{\vp}{\vec{\phi}(t,{\rm x})}
\newcommand{\vG}{\vec{G}(t,{\rm x})}
\newcommand{\ve}{\vec{\eta}(t,{\rm x})}
\newcommand{\va}{\vec{a}(t)}
\newcommand{\vc}{\vec{c}(t)}
\newcommand{\ves}{\vec{\eta}^*(t,\x)}
\newcommand{\vas}{\vec{a}^*(t)}
\newcommand{\vcs}{\vec{c}^*(t)}
\newcommand{\vA}{\vec{A}(t)}
\newcommand{\vC}{\vec{C}(t)}
\newcommand{\vGs}{\vec{G}^*(t,\x)}
\newcommand{\vAs}{\vec{A}^*(t)}
\newcommand{\vCs}{\vec{C}^*(t)}
\newcommand{\Sev}
{S_{ext}[\vec{\eta},\vec{a},\vec{c};\vec{\eta}^*,\vec{a}^*,\vec{c}^*]}
\newcommand{\LD}{{\cal L}}
\newcommand{\R}{{\rm I\!R}}

\begin{document}
\thispagestyle{empty}
{\hfill KL--TH 96/8}
\vspace{0.6cm}
\begin{center}
{\Large\bf Batalin--Vilkovisky field--antifield
quantisation of fluctuations around classical
field configurations}
\\[25mm]
F. Zimmerschied
\\[1cm]
{\it Department of Physics, \\
University of Kaiserslautern, P.\ O.\ Box 3049, D 67653 Kaiserslautern,
Germany \\
E--mail:
zimmers@physik.uni--kl.de} \\[30mm]
{\bf Abstract}
\end{center}
The Lagrangian field--antifield formalism of Batalin and Vilkovisky (BV) 
is used to investigate the application of the collective coordinate method to
soliton quantisation. In field theories with soliton solutions, the Gaussian 
fluctuation operator has zero modes due to the breakdown of global symmetries 
of the Lagrangian in the soliton solutions. It is shown how Noether identities
and local symmetries of the Lagrangian arise when collective coordinates are 
introduced in order to avoid divergences related to these zero modes. This
transformation to collective and fluctuation degrees of freedom is interpreted
as a canonical transformation in the symplectic field--antifield space which 
induces a time--local gauge symmetry. Separating the corresponding Lagrangian 
path integral of the BV scheme in lowest order into harmonic quantum 
fluctuations and a free motion of the collective coordinate with the classical 
mass of the soliton, we show how the BV approach clarifies the relation between zero modes, collective coordinates, gauge invariance and the center--of--mass 
motion of classical solutions in quantum fields. Finally, we apply
the procedure to the reduced nonlinear $O(3)$ $\sigma$--model.  
\newpage

\section{Introduction}
One of the most powerful quantisation procedures for gauge theories involving 
BRST invariance is the Lagrangian field--antifield formalism of Batalin and
Vilkovisky (BV) \cite{1,2}. The main idea is to start from the original 
Lagrangian, double the number of fields by introducing antifields of opposite
parity and then to construct an ``extended action'' as a bosonic functional
in fields and antifields which contains all the information about the dynamics 
and local symmetries of the original problem.

Together with a special bilinear form, the ``antibracket'', the fields and
antifields generate a symplectic structure which allows the application of
well--known symplectic techniques like canonical transformations. Gauge fixing,
e.\ g.\ , turns out to be a special canonical transformation \cite{2}. 
Quantization is finally performed in terms of Lagrangian path integrals over 
the gauge fixed extended action with the unphysical antifields set to zero.

Besides the standard applications of BV quantisation to theories like 
Yang--Mills field theory or general relativity (which obviously possess gauge 
symmetry and thus are constrained systems), one can also apply the BV formalism
to constrained systems which possess a hidden gauge invariance, a ``gauge 
symmetry without gauge fields'' \cite{3}, like the example of quantizing a 
particle around a classical orbit which may be done in a BRST invariant way 
\cite{4}. More 
prominent examples (which are nonetheless closely related to the classical 
orbit model) are field theories with classical, localized solutions, such as
e.\ g.\ 
solitons (or sphalerons and bounces, collectively described as solitons here).

``Quantizing solitons'', i.\ e.\ constructing quantum states of fields in
the background of the soliton 
(instead of the usual field quantum states based on 
the perturbation theory
vacuum) leads to divergences due to zero modes of the Gaussian fluctuation
operator which arise because the classical solutions break a global symmetry
of the original Lagrangian. In the past 20 years, a lot of work was done to
solve and understand this problem \cite{5,6,7,8,9}, resulting in the so--called
collective coordinate method which may be understood as a parameter dependent
transformation to new (fluctuation) fields. Interpreting the transformation 
parameters as new dynamical degrees of freedom, the configuration space 
variables of the transformed theory (i.\ e.\ collective coordinates and
fluctuations) are overcomplete so that
the resulting new Lagrangian is invariant under
local transformations and the theory has to be treated as a gauge theory.

The well--known path--integral methods of quantizing a gauge theory, 
including the choice of gauge fixing, then lead to soliton quantisation 
in the collective coordinate method. Usually Hamiltonian path integrals
are used \cite{10,11}, and the powerful tools of the BRST symmetry \cite{12}
in the Hamiltonian Batalin--Fradkin--Vilkovisky (BFV) scheme \cite{15} can
be applied to solitons \cite{7,8}. Particularly in theories with a field 
dependent
mass such as Skyrme--like models \cite{13}, it is essential to use the 
Hamiltonian
path integral since here the integration over the field momenta yields an extra
measure factor \cite{14}, a fact emphasized already 30 years ago \cite{11}. 

Since field theories start form the (Lorentz invariant) Lagrangian
formalism, it is an intriguing task to develop 
a soliton quantisation procedure which
is based on Lagrangian path integrals, avoiding completely the introduction
of field momenta. The BV scheme yields precisely 
such a Lagrangian BRST path integral
quantisation which is equivalent to the BFV method \cite{15a}, so that the
measure problem referred to 
in Lagrangian path integrals should not arise in BV Lagrangian
path integrals (although theories with non--trivial measures are also
discussed in the context of the BV scheme \cite{15b}).

Formally it
is straightforward to apply this Lagrangian BV scheme to the quantisation
of solitons \cite{9}. We do this in the following by application to 
a simple field theory model with the intent to 
study in a transparent way
the pecularities of the symplectic field--antifield concept. 
Thereby, our approach to the method of 
collective coordinates is different from that in \cite{9},
as we introduce collective coordinates and fluctuations together within one
and the same
transformation from the beginning to the new, overcomplete set of fields. In 
the field--antifield formalism, this transformation may be understood as a
canonical transformation from the old theory (with global symmetries
resulting in zero modes of the Gaussian fluctuation operator) to the
new theory with local gauge symmetries which are given by the canonical
transformation of the extended action in BV.

Then analysing the Lagrangian path integral, we show that the BV scheme
emphasises the semiclassical interpretation of solitons as localised particles 
in Lagrangian field theories: Neglecting quantum corrections to the classical 
soliton mass in lowest order, we may separate the center--of--mass motion of
the soliton (described by its collective coordinate) from the quantum 
fluctuations around it in the harmonic or one--loop approximation.

Finally, we give some hints concerning the application of  
the BV scheme to Mottola and Wipf's reduced $O(3)$ 
$\sigma$--model \cite{16} in a spherical field parametrisation which yields a 
Lagrangian of the general form considered in \cite{9}. Besides collective
translation degrees of freedom which we discussed so far, collective rotation 
degrees in the internal symmetry space arise in this model which in the context
of the Hamiltonian BFV technique and for a slightly different $\sigma$--model
were discussed in \cite{16a}.

The main advantage of BV quantisation, from the point of view of soliton 
physics, therefore is not its capability to handle complicated gauge structures
like open gauge algebras (which was one of the original aims of Batalin and 
Vilkovisky), but its simple and straight--forward application to collective 
coordinates and fluctuations which yields a general procedure and formal 
structures with clear physical interpretation.
The structures of the gauge algebras discussed here are quite
simple; no open gauge algebras appear. This allows the investigation of 
some special topics of the BV formalism such 
as the relation between infinitesimal 
canonical transformations, gauge fixing and BRST transformations in the 
context of a simple example in a more transparent way than in the context of 
abstract considerations.

\section{Collective coordinates, singular properties of the fluctuation
Lagrangian and the field--antifield formalism}
\label{section1}

We consider the scalar field theory of a bosonic ($p(\phi)=0$) or fermionic
($p(\phi)=1$, $p$ denoting the Grassmann parity) field $\phi(t,\x)$ in
$1+1$ dimensions with Lagrangian density
\begin{equation}
\tilde{\LD}(\phi,\phi';\dot{\phi}) = \frac12(\dot{\phi}^2-\phi^{\prime 2}) - 
V(\phi).
\label{1}
\end{equation}
Dots denote derivatives with respect to time $t$, primes with respect to space
$\x$. 
We assume that this model contains a soliton configuration
$\varphi(\x)$, i.\ e.\ a nontrivial,
classical, static solution of finite energy (most prominent examples being
$\phi^4$ and Sine--Gordon theories).
The Lagrangian has a global space translation invariance $\x\mapsto \x-a$,
$a\in\R$ which is broken by the soliton solution as $\varphi(\x-a)$ is a new
solution different from $\varphi(\x)$. The parameter 
$a$ describes the localisation of the
soliton.

We now use the classical solution $\varphi$ to introduce new ``fields'' 
$\eta(t,\x)$ and $a(t)$ (which in fact is not a field, but a 
dynamical parameter) by setting
\begin{equation}
\phi(t,\x) = \varphi(\x-a(t))+\eta(t,\x)
\label{2}
\end{equation}
which means that we add one more
degree of freedom (the ``collective coordinate''
$a(t)$) to the infinitely many field degrees of freedom. We denote
these new ``fields'' collectively
by the symbol $\Phi^i(x)\in\{\eta(t,\x),a(t)\}$;
$x$ denoting either spacetime or time. 

The transformed Lagrangian density then reads
\begin{eqnarray}
\LD(\x;\eta,a,\eta';\dot{a},\dot{\eta}) 
& = & \frac12\left[-\varphi'(\x-a(t))\dot{a}(t)+\dot{\eta}(t,\x)\right]^2
\nonumber \\
& & {} -\frac12\left[-\varphi'(\x-a(t))+\eta'(t,\x)\right]^2
\nonumber \\
& & {} -V(\varphi(\x-a(t))+\eta(t,\x))
\label{3}
\end{eqnarray}

This new Lagrangian density
$\LD$ depends explicitly on the space coordinate $\x$. This is a first hint to 
a special feature of collective coordinate techniques: In our model, we have 
a mixture of 
fields (the fluctuations $\eta(t,\x))$) and coordinates (the collective 
coordinate $a(t)$ describing a translation in space). The collective coordinate
itself does not depend on space: It is an additionel and redundant degree of 
freedom to the infinitely many field degrees of freedom labeled by the 
continuous index ``space''. In this sense, one may interpret the fluctuation
field $\eta(t,\x)$ as an 
infinite set of ``fluctuation coordinates'' which could be
made more explicit by use of the notation 
$\eta(t,\x)=a_{\x}(t)$. From this point of
view, the collective coordinate method is more closely related  
to Lagrangian mechanics than
to Lagrangian field theory. In particular, our model may be regarded as
the $N\longrightarrow\infty$ limit of the model of a particle around 
an orbit in $N$ dimensions \cite{3,4}.

This special role of space position 
arises again when we evaluate the Euler--Lagrange
equations of our model. As usual, the dynamics is determined by extremizing 
the full
action
\begin{equation}
S_0[\eta,a] = \int \LD(\x;\eta,a,\eta';\dot{a},\dot{\eta}) d\x dt
\label{4}
\end{equation}
with respect to $\eta$ and $a$. Therefore, setting the corresponding functional
derivatives to zero,
\begin{equation}
\left(S_0[\eta,a]\frac{\dr}{\delta\eta(t,\x)}\right) \stackrel{!}{=} 0, \qquad
\left(S_0[\eta,a]\frac{\dr}{\delta a(t)}\right) \stackrel{!}{=} 0
\label{5} \label{6}
\end{equation}
leads to the Euler--Lagrange equations. (In the BV formalism, we have to 
distinguish between derivatives from the right and from the left even when 
starting with a bosonic theory: The introduction of antifields later  
unavoidably leads on to fermionic degrees of freedom. Using
derivatives from the right in eq.\ (\ref{6})
is more or less a convention which has to
be consistent with all later formulae.)

We see that the Euler--Lagrange 
equation of the collective coordinate does not depend on the position space
coordinate $\x$ which remains
integrated out:
\begin{equation}
\left(S_0[\eta,a]\frac{\dr}{\delta a(t)}\right) = 
\int\left[\frac{\partial\LD}{\partial a} - 
\frac{d}{dt}\frac{\partial \LD}{\partial\dot{a}}\right]d\x.
\label{8}
\end{equation}

The two equations (\ref{5}) are not 
independent: As a trivial consequence of adding one degree of freedom,
the relation
\begin{equation}
\int \left(S_0[\eta,a]\frac{\dr}{\delta\eta(t,\x)}\right)\cdot
\phi'(\x-a(t)) d\x + \left(S_0[\eta,a]\frac{\dr}{\delta a(t)}\right) \cdot 1
=0
\label{9}
\end{equation}
obtained by inspection of the explicit forms of eqs. (\ref{5})
holds identically. According to the above discussion, we may interpret the
integral in eq.\ (\ref{9}) as a sum over the continuous index $\x$ of the
``fluctuation coordinates''. Then (\ref{9}) is an ordinary Noether identity
\cite{17} of the form
\begin{equation}
\left(S_0[\Phi]\frac{\dr}{\delta \Phi^i(x)}\right) R^i_{\alpha}[\Phi](x)
\equiv 0,
\quad \alpha=1,\ldots,m
\label{10}
\end{equation}
($\alpha=m=1$ in our case) with ordinary Noether generators 
$R^i_{\alpha}[\Phi](x)$. These Noether identities ensure that local 
$m$--parameter symmetries
\begin{equation}
\delta_{(\epsilon)}\Phi^i(x) = \epsilon^{\alpha}(x) R^i_{\alpha}[\Phi](x)
\label{11}
\end{equation}
hold as can be seen by Taylor expansion of the transformed action
$S_0[\Phi+\delta_{(\epsilon)}\Phi]$.

Noether identities of the form (\ref{10}) are called ``ordinary'' to distinguish
them from generalised Noether identities
\begin{equation}
\int\left(S_0[\Phi]\frac{\dr}{\delta \Phi^i(x')}\right) R^i_{\alpha}[\Phi](x,x')
dx'\equiv 0
\label{12}
\end{equation}
with related local symmetries ($R^i_{\alpha}[\Phi](x,x')\propto\delta(x-x')$)
\begin{eqnarray}
\delta_{(\epsilon)}\Phi^i(x) & = & \int \epsilon^{\alpha}(x') 
R^i_{\alpha}[\Phi](x,x')dx' \nonumber \\
& = & \epsilon^{\alpha}(x) {}^{(0)}\!R^i_{\alpha}[\Phi](x) +
\partial_{\mu}\epsilon^{\alpha}(x) {}^{(1)}\!R^{i\mu}_{\alpha}[\Phi](x)
\label{13}
\end{eqnarray}
depending on local parameters $\epsilon^{\alpha}(x)$ and their derivatives
\cite{2,17}. These are typical symmetries of gauge fields. The fact that
only ordinary Noether identities and thus gauge transformations of the form
(\ref{11}) arise in our model justifies the expression ``gauge theory
without gauge fields'' \cite{3}: Although invariant under local transformations,
there are no fields which transform like gauge fields in our model.

By comparison of (\ref{9}) and (\ref{10}), we can 
identify the Noether generators of our model,
\begin{eqnarray}
R^{(\eta)}[a](t,\x) & = & \varphi'(\x-a(t))\label{14} \\
R^{(a)} & = & 1
\label{15}
\end{eqnarray}
Again, the position space coordinate in (\ref{14}) should be considered as a
continuous
index: The fact that it is integrated out in (\ref{9}) yields that the related
gauge symmetry is only local in time:
\begin{eqnarray}
\delta_{(\epsilon)} \eta(t,\x) & = & \epsilon(t) \varphi'(\x-a(t)) \label{16} \\
\delta_{(\epsilon)} a(t) & = & \epsilon(t) \label{17}
\end{eqnarray}
or in more general notation
\begin{equation}
\delta_{(\epsilon)} \Phi^i(x) = \epsilon^{\alpha}(t)R^i_{\alpha}[\Phi](x)
\qquad (\alpha=1).
\label{18}
\end{equation}

Eq. (\ref{16},\ref{17}) 
are typical time--local gauge transformations associated with
collective translation coordinates: They depend on $\frac{d}{dx}\varphi$, 
where $\frac{d}{dx}$ is the generator of translations ($\varphi'$ is also the
zero mode of the Gaussian fluctuation operator of the model as we shall 
discuss later). We observe later 
that collective rotation coordinates (which
arise e.\ g. in $\sigma$--models) also yield typical, time--local gauge
transformations related to some generator of rotations.

Certainly, the gauge algebra desribed by (\ref{16},\ref{17}) is trivially
abelian. But also in models with more than one collective coordinate
(which have more than one Noether identity), the gauge algebras are closed
and often abelian (in particular, this is true for $\sigma$--models).

In accordance with the usual BRST treatment of gauge theories, we now introduce 
one ghost ``field'' (which in fact is a coordinate as it depends only on time)
$c(t)$ with parity opposite to that of the symmetry parameter $\epsilon(t)$:
$p(c)=p(\epsilon)+1$. From (\ref{17}), $p(\epsilon)=p(a)$, and as space has
bosonic properties in our model, the ghost $c(t)$ is fermionic. The 
configuration space now consists of $\{\eta(t,\x),a(t),c(t)\}=:\{Y^A(x)\}$
where we use $Y^a(x)$ as abbreviation for any field in the extended field
configuration spaces which we construct in the following (whereas $\Phi^i(x)$ 
denotes only the original fields, i.\ e.\ $\eta(t,\x)$ and $a(t)$ in our 
example).

From the local symmetry (\ref{18}), we now construct the global BRST symmetry
\begin{equation}
\delta_{\lambda}Y^A(x) = \lambda\s Y^A(x)
\label{19}
\end{equation}
where $\s$ is the BRST operator which is defined by the following properties:
(1) $\s$ is parity changing: $p(\s Y^A(x))=p(Y^A(x))+1$, (2) $\s$ is nilpotent,
$\s^2=0$, (3) $\s$ ``includes'' the gauge symmetry in the sense that $S_0[\Phi]$
is invariant under $\delta_{\lambda}\Phi^i(x)=\lambda\s \Phi^i(x)$.

From these properties and (\ref{11},\ref{12}) 
it is easy to conclude that the BRST
operator of our model is given by
\begin{eqnarray}
\s \eta(t,\x) & = & R^{(\eta)}[a](t,\x)c(t) = \varphi'(\x-a(t))c(t) \label{20}\\
\s a(t) & = & R^{(a)}c(t) = c(t) \label{21} \\
\s c(t) & = & 0 \label{22}
\end{eqnarray}

Next, we double the number of degrees of freedom of the ghost--enlarged
configuration space by introducing ``antifields'' of opposite parity. Denoting
antifields by an asterix, we thus have the ``fields'' 
$\{Y^A(x)\}=\{\eta(t,\x),a(t),c(t)\}$ and the ``antifields''
$\{Y^*_A(x)\}=\{\eta^*(t,\x),a^*(t),c^*(t)\}$ with $p(Y^*_A(x))=p(Y^A(x))+1$.
Together with the antibracket of two functionals $U$, $V$ defined by
\begin{equation}
(U,V) = \int\left\{\left(U\frac{\dr}{\df}\right)\left(\frac{\dl}{\da}V\right) -
\left(U\frac{\dr}{\da}\right)\left(\frac{\dl}{\df}V\right)\right\}dx,
\label{23}
\end{equation}
the configuration space 
\begin{equation}
{\cal P}=\{Y^A(x),Y^*_A(x)\}=
\{\eta(t,\x),a(t),c(t);\eta^*(t,\x),a^*(t),c^*(t)\}
\label{fa}
\end{equation}
has a symplectic structure similar to that of Hamiltonian phase space with the
Poisson bracket. But in contrast to the Poisson bracket, $(U,U)=0$ holds only 
for fermionic functionals $U$ with 
$p(U)=1$, whereas for bosonic functionals $S$ with $p(S)=0$
\begin{equation}
(S,S) = 2\int\left(S\frac{\dr}{\df}\right)\left(\frac{\dl}{\da}S\right).
\label{24}
\end{equation}
If the nontrivial equation $(S,S)=0$ is fullfilled for some 
bosonic $S$, then the
identity
\begin{eqnarray}
0 & = & (S,S)\frac{\dr}{\delta Y^*_B(x)} = 
\left(S,S\frac{\dr}{\delta Y^*_B(x)}\right) \nonumber\\
& = & 2 \int\left\{\left(S\frac{\dr}{\delta Y^A(x')}\right) 
\left(\frac{\dl}{\delta Y^*_A(x')}S\frac{\dr}{\delta Y^B(x)}\right)\right.
\nonumber \\
& & \qquad\qquad \left.
{} - \left(S\frac{\dr}{\delta Y^*_A(x')}\right)
\left(\frac{\dl}{\delta Y^A(x')}S\frac{\dr}{\delta Y^*_B(x)}\right)\right\} dx'
\label{25}
\end{eqnarray}
and a similar one for $(S,S)\frac{\dr}{\delta Y^B(x)}$ holds. Eq. (\ref{25})
resembles the general form of the Noether identity (\ref{10}) if
$S\frac{\dr}{\delta Y^A(x')}$ is related to the Euler--Lagrange equations (here
we see that the directions of the derivatives in (\ref{5}) and (\ref{23}) were
chosen consistently).
This purely structural feature of bosonic functionals on the symplectic 
field--antifield space (\ref{fa}) is exploited physicswise 
in the construction of
the socalled ``extended action'' 
$\Se$. This extended action is a bosonic functional which describes 
both the dynamics
and the gauge symmetries of a given theory. It is defined as solution of the  
``classical master equation''
\begin{equation}
\left(\Se,\Se\right) = 0
\label{26}
\end{equation}
with boundary conditions 
\begin{equation}
\Se|_{Y^*=0}=S_0[\Phi]
\label{first}
\end{equation}
and 
\begin{equation}
\left.\frac{\dl}{\delta c^{\alpha}(x)}\Se\frac{\dr}{\delta \Phi^*_i(x)}
\right|_{Y^*=0} = R^i_{\alpha}[\Phi](x)
\label{second}
\end{equation}
where $c^{\alpha}$ is the general notation for the ghost fields. 

The first
boundary condition (\ref{first})
retains the dynamics of the model: In the BV scheme, the antifields have no
physical meaning --- to obtain physical quantities, they are set to zero (a fact
which becomes important in the evaluation of path integrals). The ``physical
part'', in this sense, of the extended action is thus the original action
which describes the dynamics of the theory.
The second boundary condition (\ref{second}) uses the relation (\ref{25})
to enforce the local symmetries of the theory (described by the Noether 
identities (\ref{10}))
into the extended action. In a naive way, on could say that the redundant 
degrees of freedom we gained by introducing antifields enabled
us to ``add'' to the action 
the important information about the symmetry structure
which is contained in the Noether generators, resulting in the ``extended 
action''.

The solution of the classical master equation is a crucial step in the BV
scheme. Eq.\ (\ref{first}) suggests an ansatz in powers of antifields,
\begin{equation}
\Se = \sum_{n=0}^{\infty} \int Y^*_{A_n}(x_n)\cdot\ldots\cdot Y^*_{A_1}(x_1)
S^{A_1\ldots A_n}[Y](x_1,\ldots, x_n) dx_1\ldots dx_n
\label{ansatz}
\end{equation}
For abelian gauge theories, the structure constants of the
gauge algebra vanish, and one can show that no terms higher than those of
linear order are necessary in (\ref{ansatz}) to solve the master equation
(this corresponds to the fact that the BRST transformations of the ghost
variables are trivial, $\s c^{\alpha}(x)=0$, in these cases). In particular,
we have from (\ref{second}) that
\begin{equation}
\frac{\dl}{\da}\Se = \s Y^A(x)
\label{27}
\end{equation}
($Y^*=0$ is no longer necessary!), and the extended action is simply given by
\begin{equation}
\Se=S_0[\Phi] + \int Y^*_A(x)\left(\s Y^A(x)\right)dx.
\label{28}
\end{equation}
With (\ref{fa}) and (\ref{20},\ref{21},\ref{22}), this implies
for our soliton model
\begin{equation}
\Sext = 
S_0[\eta,a] + \int a^*(t)c(t)dt + \int \eta^*(t,\x)\varphi'(\x-a(t))c(t)dtd\x.
\label{29}
\end{equation}

From (\ref{27}) and the definition of the antibracket (\ref{23}), we see that 
in the field--antifield formalism, the extended action may be regarded as the
analogue of the BRST charge in the Hamiltonian BFV formalism \cite{15} in the 
sense that it generates the BRST transformations:
\begin{equation}
\s Y^A(x) = \left(Y^A(x),\Se\right)
\label{30}
\end{equation}
Eq.\ (\ref{30}) also holds for nonabelian gauge algebras with more complicated
BRST transformations than (19). 

It is now obvious how we have to define the BRST transformations of the 
antifields which have not been defined so far: We simply set
\begin{equation}
\s Y^*_A(x) = \left(Y^*_A(x),\Se\right).
\label{31}
\end{equation}
With this definition, $\s$ is nilpotent on all functionals over ${\cal P}$, and
the classical master equation ensures that the extended action is BRST 
invariant:
\begin{equation}
\s\Se = \left(\Se,\Se\right) = 0
\label{32}
\end{equation}

\section{Canonical field--antifield transformations and collective coordinates}

The construction of the extended action demonstrated 
how purely structural properties
(the fact that $(S,S)=0$ is a nontrivial equation for bosonic $S$) of the 
symplectic field--antifield space can be used to formulate physical aspects of 
the theory as dynamics and symmetry by the construction of the extended action
which is the main object of the BV formalism. 

Another important instrument in 
the study of symplectic structure is the group of 
transformations which leave the symplectic form invariant, i.\ e.\ the
antibracket in the 
BV scheme. In classical mechanics, these symmetries are called ``canonical
transformations''. As in Hamiltonian mechanics, they are usually defined via 
a generator which in the BV case has to be a fermionic functional of e.\ g.\
old fields $\tilde{Y}^A(x)$ and new antifields $Y^*_A(x)$, 
$F[\tilde{Y},Y^*]$. The canonical transformation 
$\{\tilde{Y}^A(x),\tilde{Y}^*_A(x)\}\mapsto \{Y^A(x),Y^*_A(x)\}$ 
is then given by
\cite{2}
\begin{equation}
\tilde{Y}^*_A(x) = \frac{\dl}{\delta \tilde{Y}^A(x)}F[\tilde{Y},Y^*], \qquad
Y^A(x) = \frac{\dl}{\da}F[\tilde{Y},Y^*].
\label{33}
\end{equation}
Since $F$ is fermionic, canonical transformations do not change parity.

We now use canonical transformations to introduce collective coordinates and
fluctuations, starting from the original Lagrangian (\ref{1}) and its global
symmetry $\x\mapsto\x-a$ which is broken by the soltion solution. For this
(gobal!) symmetry, we add one absolutely redundant bosonic symmetry coordinate
$A(t)$ --- in general, one has to add one such coordinate for each symmetry
degree of freedom with corresponding parity.

Since $\tilde{\LD}(\phi,\phi';\dot{\phi})$ does not depend on $A(t)$ at all, we
have $\tilde{S}_0[\phi,A]\frac{\dr}{\delta A(t)}=0$ where
$\tilde{S}_0[\phi,A]=\int\tilde{\LD}(\phi,\phi';\dot{\phi})dtd\x$, and a
trivial Noether identity
\begin{equation}
\int \left(\tilde{S}_0[\phi,A]\frac{\dr}{\delta \phi(t,\x)}\right)R^{(\phi)} d\x
+ \left(\tilde{S}_0[\phi,A]\frac{\dr}{\delta A(t)}\right)R^{(A)} = 0
\label{34}
\end{equation}
with $R^{(\phi)}=0$, $R^{(A)}=1$ holds. In order to obtain the BRST 
transformations associated with
(\ref{34}), we introduce one fermionic ghost
coordinate $C(t)$. Then $\tilde{S}_0[\phi,A]$ is trivially BRST invariant under
\begin{eqnarray}
\s \phi(t,\x) & = & R^{(\phi)}C(t) = 0 \nonumber \\
\s A(t) & = & R^{(A)}C(t) = C(t) \nonumber \\
\s C(t) & = & 0.
\label{35}
\end{eqnarray}

Doubling the (old) fields $\{\tilde{Y}^A(x)\}=\{\phi(t,\x),A(t),C(t)\}$ by
introducing
antifields $\{\tilde{Y}^*_A(x)\}=\{\phi^*(t,\x),A^*(t),C^*(t)\}$, we
can immediately write down the extended action
\begin{equation}
\tilde{S}_{ext}[\phi,A,C;\phi^*,A^*,C^*] = \tilde{S}_0[\phi,A]
+ \int A^*(t)C(t)dt.
\label{36}
\end{equation}

We exploit the fact that due to the global symmetry of (\ref{1}) and the 
existence of a soliton solution $\varphi(\x)$, we have a whole class of
solutions $\varphi(\x-a)$, $a\in\R$. Using these we write out the
fermionic generator
\begin{eqnarray}
& & 
F[\phi,A,C;\eta^*,a^*,c^*] \nonumber \\
& = & \int a^*(t)A(t)dt + \int c^*(t)C(t)dt
+ \int \eta^*(t,\x)[\phi(t,\x)-\varphi(\x-A(t)] dtd\x
\label{37}
\end{eqnarray}
which depends on old fields and new antifields and yields the canonical 
transformation
\begin{eqnarray}
\phi^*(t,\x) & = & \frac{\dl}{\delta \phi(t,\x)}F[\phi,A,C;\eta^*,a^*,c^*]
= \eta^*(t,\x) \nonumber \\
A^*(t) & = & \frac{\dl}{\delta A(t)}F[\phi,A,C;\eta^*,a^*,c^*]
= a^*(t)+ \int\eta^*(t,\x)\varphi'(\x-A(t))d\x\nonumber \\
C^*(t) & = & \frac{\dl}{\delta C(t)}F[\phi,A,C;\eta^*,a^*,c^*]
= c^*(t)\nonumber \\
\eta(t,\x) & = & \frac{\dl}{\delta \eta^*(t,\x)}F[\phi,A,C;\eta^*,a^*,c^*]
= \phi(t,\x) - \varphi(\x-A(t)) \nonumber \\
a(t) & = & \frac{\dl}{\delta a^*(t)}F[\phi,A,C;\eta^*,a^*,c^*]
= A(t) \nonumber \\
c(t) & = & \frac{\dl}{\delta c^*(t)}F[\phi,A,C;\eta^*,a^*,c^*] = C(t).
\label{39}
\end{eqnarray}
We thus have (as in (\ref{2})) $\phi(t,\x)=\varphi(\x-a(t))+\eta(t,\x)$. 
Inserting (\ref{39}) into the extendend action (\ref{39}) yields
\begin{eqnarray}
& & \tilde{S}_{ext}[\phi,A,C;\phi^*,A^*,C^*] \stackrel{F}{\mapsto}
\Sext \nonumber \\
& & \qquad {}= 
S_0[\eta,a] + \int a^*(t)c(t)dt + \int \eta^*(t,\x)\varphi'(\x-a(t))c(t)dtd\x,
\label{39a}
\end{eqnarray}
and from the general form of the extended action in abelian theories,
i.\ e.\ (\ref{28}),
we can extract from (\ref{39a}) the BRST transformations (\ref{20}--\ref{22}) of
the classical action in collective and fluctuation degrees of freedom,
$S_0[\eta,a]$. We thus obtain the gauge symmetry together with the collective--
fluctuation action by a canonical transformation.

The crucial point of this procedure is the construction of the generator 
(\ref{37}) which has a simple physical interpretation: The completely
redundant coordinate $A(t)$ is correlated to the symmetry parameter $a$ of
the original Lagrangian which becomes dynamical, and the difference between 
the classical soliton solution $\phi$ at $A(t)=a(t)$
and the original field $\phi(t,\x)$ defines the fluctuations $\eta(t,\x)$, a definition similar to that
of center--of--mass frame coordinates in Lagrangian mechanics and related 
to the model of the classical orbit \cite{4,19}. 

The approach of eq.\ (\ref{2}) to define fluctuations which we gained
here by a canonical transformation is commonly used in
the collective coordinate method \cite{5}, 
but there is also a different classical
way to introduce collective coordinates \cite{6} which starts from
\begin{equation}
\phi(t,\x) = \tilde{\phi}(t,\x-a(t))
\label{40}
\end{equation}
thus using $\{\tilde{\phi},a\}$ as a new and 
overcomplete set of configuration space 
variables. Proceeding along these lines, fluctuations are introduced later 
by expanding $\tilde{\phi}(t,\rho)$ with $\rho=\x-a(t)$ around $\varphi(\rho)$,
where $\varphi$ again is the soliton solution. This yields
\begin{equation}
\phi(t,\x)=\varphi(\x-a(t)) + \eta(t,\x-a(t))
\label{41}
\end{equation}
which is different from (\ref{2}) since the fluctuations now also depend on
the collective coordinate shifted space argument. This may be interpreted
as a common frame of reference in which both the soliton and the fluctuations
(frequently visualised as analogues of 
baryon and mesons) are observed. This point of view 
is useful in the discussion of soliton--fluctuation scattering, but one looses
the interpretation of the space coordinate as a continuous index labeling the
``fluctuation coordinates'' $a_{\x}(t)=\eta(t,\x)$ which arises directly
from the integral form of the Noether identity (\ref{9}). 

Thus in the BV scheme, the introduction of collective coordinates and 
fluctuations according to (\ref{2}), which is related to a canonical 
transformation in field--antifield space, is more convenient.

Another important application of canonical transformations in the BV scheme is
the gauge fixing of the theory which we discuss in the next section.
Besides canonical transformations (which are the symmetry transformations
of the symplectic structure of the field--antifield space), there is still 
the BRST symmetry as a symmetry of the extended action. The BRST transformation
is not canonical since it changes parity. There is, however, a close relation
between the BRST transformation and canonical transformations. To see this,
we have to analyse parameter groups of canonical transformations generated
by $F_{\alpha}[\tilde{Y},Y^*]$ ($\alpha$ being the global parameter of
the transformation group). As usual, on can introduce infinitesimal canonical 
transformations generated by $f[Y,Y^*]$ where $f$ is the linear coefficient
in the $\alpha$--expansion of $F_{\alpha}[\tilde{Y},Y^*]$ (replacing $\tilde{Y}$
by $Y$):
\begin{eqnarray}
\delta_{\alpha} Y^A(x) & = & \alpha (Y^A(x),f[Y,Y^*]) \nonumber \\
\delta_{\alpha} Y^*_A(x) & = & \alpha (Y^*_A(x),f[Y,Y^*]). 
\label{42}
\end{eqnarray}
The infinitesimal canonical transformation of a genral functional $U$ is then
given by $\delta_{\alpha}U = \alpha(U,f[Y,Y^*])$. Introducing some condensed
notation, we may assign an operator ${\bf X}_f$ to the generating functional
$f[Y,Y^*]$, defined by ${\bf X}_fA:=(A,f[Y,Y^*])$. Using this notation 
naively, we have $\s={\bf X}_{S_{ext}}$ for the BRST operator: But as $\Sext$ is
bosonic, ${\bf X}_{S_{ext}}$ changes parity. Parity--changing transformtions in
general do not leave the antibracket invariant and thus cannot be canonical.

\section{Gauge fixing and path integrals}
\label{section3}

Before evaluating path integrals to quantize our model, we have to gauge--fix
the extended action (\ref{29}). To do this, it is necessary to add a ``trivial
system'' for each Noether identity (\ref{10}) of the theory and so to enlarge 
the symplectic field--antifield space again. 

There is only one Noether identity (\ref{9}) in our model, so one trivial
system $\bar{c}(t)$, $b(t)$ with $\s\bar{c}(t)=b(t)$, $\s b(t) = 0$, and
the corresponding antifields $\bar{c}^*(t)$, $b^*(t)$ have to be added.
$\bar{c}$ and $b$ have opposite parity (which is obvious from their BRST
transformations), and $p(\bar{c})=p(c)$. We ``add'' the BRST transformation
of the trivial system to the extended action, yielding
\begin{eqnarray}
& & \Sextt \nonumber \\
& = & 
S_0[\eta,a] + \int a^*(t)c(t)dt + \int \eta^*(t,\x)\varphi'(\x-a(t))c(t)dtd\x
+\int\bar{c}^*(t)b(t)dt. \nonumber \\
\label{43}
\end{eqnarray}
The trivial system is needed to construct the ``gauge fermion'', a fermionic
functional $\psi[Y]$ which depends only on fields and not on antifields
($Y$ again denotes all fields, i.\ e.\ now including the trivial system). Since 
$\psi$ is fermionic, it may be regarded as generator of an infinitesimal
canonical transformation. Gauge fixing the extended action then simply
means performing this transformation:
\begin{equation}
\Se \stackrel{\psi}{\mapsto} \Sep = \Se + \alpha {\bf X}_{\psi}\Se
\label{44}
\end{equation}
where $\Sep$ denotes the $\psi$--gauge fixed extended action. Here, we use 
infinitesimal canonical transformations to perform the gauge fixing and will 
set the transformation parameter $\alpha$ to 1 later. It is easy to see that
one can also use canonical transformations to perform the gauge fixing
\cite{2}, but the relation between BRST transformations (which are also
given in infinitesimal form) and gauge fixing becomes clearer when 
infinitesimal canonical transformations are used to gauge fix the extended
action.

This relation is due to the fact that the gauge fermion $\psi[Y]$ generates
an infinitesimal canonical transformation, whereas the extended action
generates the BRST transformation. So
\begin{equation}
{\rm X}_{\psi}\Se = \left(\Se,\psi[Y]\right) = - \left(\psi[Y],\Se\right)
= - \s\psi[Y].
\label{45}
\end{equation}
Gauge fixing, from this point of view, consists of adding the BRST--transformed
gauge fermion to the extended action, a term which is trivially BRST invariant
due to the nilpotency of $\s$. This is very similar to the gauge fixing 
procedure in the Hamiltonian BFV scheme \cite{3,15}, and the trivial system 
fields $\bar{c}(t)$, $b(t)$ are comparable to the ``antighost'' and
the Nakanashi--Lautrup field in BFV (``antighosts'' in BFV are not related
to the antifields of the ghosts in BV!).

The main task of all gauge fixing procedures is to check that quantized
physical quantities (e.\ g.\ expectation values defined by path integrals) 
do not depend on the choice of the specific gauge. (In fact, the proof of this 
independence is the main subject of the classical BFV papers \cite{15}).
In the BV formalism, this means that path integrals have to be invariant
under canonical transformations which is not a trivial requirement as the 
measure ${\cal D}\{Y^A(x)\}{\cal D}\{Y^*_A(x)\}$ on field--antifield space
is not invariant under canonical transformations (different from the 
Hamiltonian phase space measure for which the Liouville theorem holds).
This invariance requirement determines the right measure on the fields
$\mu({\cal D}\{Y^A(x)\})$: There is no integration over antifields in
BV-Lagrangian path integrals; as we already mentioned, 
the antifields are unphysical degrees of freedom
and have to be set to zero to obtain physical quantities.

The measure $\mu$ is usually contained the exponential: One defines
the functional $W[Y,Y^*]$ by 
$\mu({\cal D}\{Y^A(x)\})\exp\left(\frac{i}{\hbar}\Sep\right)={\cal D}\{Y^A(x)\}
\exp\left(\frac{i}{\hbar}W[Y,Y^*]\right)$.
Then one can check that the expectation value
\begin{eqnarray}
\langle\chi\rangle[Y^*] & = & \int\mu({\cal D}\{Y^A(x)\})\chi[Y,Y^*]
\exp\left(\frac{i}{\hbar}\Sep\right) \nonumber\\
& = & \int {\cal D}\{Y^A(x)\}\chi[Y,Y^*]
\exp\left(\frac{i}{\hbar}W[Y,Y^*]\right) 
\label{46}
\end{eqnarray}
of an arbitrary functional $\chi[Y,Y^*]$ (the antifields still have to
be set to zero) is invariant under infinitesimal canonical 
transformations if the relations
\begin{eqnarray}
\left(W[Y,Y^*],W[Y,Y^*]\right) - 2i\hbar\triangle W[Y,Y^*] & = & 0 \label{47}\\
\frac{i}{\hbar}\left(\chi[Y,Y^*],W[Y,Y^*]\right) + \triangle\chi[Y,Y^*] = 0
\label{48}
\end{eqnarray}
with
\begin{equation}
\triangle U[Y,Y^*] = (-1)^{p(Y^A)+p(U)}\int\frac{\dl^2}{\da\df}U[Y,Y^*]dx
\label{49}
\end{equation}
for any functional $U[Y,Y^*]$ hold. Eq.\ (\ref{47}) is the socalled 
``quantum master equation'' and (\ref{48}) defines the quantum BRST operator
\begin{equation}
\hat{\s} := \left(\cdot,W[Y,Y^*]\right)-i\hbar\triangle
\label{50}
\end{equation}
which is nilpotent if $W[Y,Y^*]$ satisfies the quantum master equation. The
cohomology classes of $\hat{\s}$ define the physical observables.

The quantum master equation is usually solved by an ansatz in powers of
$\hbar$, 
\begin{equation}
W[Y,Y^*]=\Sep+\sum_{n=1}^{\infty}\hbar^n M_n[Y,Y^*].
\label{qmeansatz}
\end{equation}
The  ${\cal O}(\hbar^0)$ order reproduces the classical master equation, and the
${\cal O}(\hbar^1)$ order correction is given by
\begin{equation}
\left(\Sep,M_1[Y,Y^*]\right) = i\triangle\Sep.
\label{51}
\end{equation}

We now apply the general procedure of gauge fixing in the BV formalism to our
soliton model. As usual, the physical idea of gauge fixing in this context
is to eliminate the fluctuations parallel to the zero mode $\varphi'$ of
the Gaussian fluctuation operator. The  corresponding gauge fermion is
\begin{equation}
\psi[\eta,a,\bar{c}] = 
\int \bar{c}(t)\left(\int\eta(t,\x)\varphi'(\x-a(t))d\x\right)dt
\label{52}
\end{equation}
where the $\x$--integral is the orthogonality condition $\eta\perp\varphi'$.
Another motivation for the choice of gauge fixing will be given later.

From the general formula (\ref{44}), we obtain the gauge--fixed extended action
\begin{eqnarray}
& & \Sexttp \nonumber \\
& = & S_0[\eta,a] + 
\int \left[a^*(t)+\bar{c}(t)\left(\int\eta(t,\x)\varphi''(\x-a(t))d\x\right)
\right]c(t)dt \nonumber \\
& & \qquad\quad {}
+ \int \int \left[ \eta^*(t,\x) -\bar{c}(t)\varphi'(\x-a(t))\right]
\varphi'(\x-a(t))c(t)d\x dt \nonumber \\
& & \qquad\quad {}
+ \int \left[\bar{c}^*(t) - \left(\int\eta(t,\x)\varphi'(\x-a(t))d\x\right)
\right]b(t)dt.
\label{53}
\end{eqnarray}
It is easy to check that $\triangle\Sexttp=0$, so there will be no quantum 
corrections to the classical master equation in our model. In more general
situations, but still with abelian gauge algebras, we have from (\ref{28})
and (\ref{45})
\begin{equation}
\Sep = S_0[\Phi] + \int Y^*_A(x)\left(\s Y^A(x)\right)dx - \s\psi[Y]
\label{54}
\end{equation}
and hence, using the fact that the gauge fermion depends only on fields,
\begin{equation}
\triangle\Sep = (-1)^{p(Y^A)}\int\frac{\dl}{\df}\left(\s Y^A(x)\right)dx.
\label{55}
\end{equation}
For a large number of collective coordinate models and especially for 
collective translation and rotation degrees of freedom as we discuss them
here, $\s Y^A(x)$ does not depend on $Y^A(x)$ itself. Otherwise, terms
proportional to $\frac{\dl}{\df}Y^A(x)\propto\delta(0)$ arise in (\ref{55}),
and an appropriate regularisation procedure is necessary.

From (\ref{55}) we see that quantum measure corrections to the BV--Lagrangian
path integral arise only from the BRST symmetry structure of the theory and
not from the action itself. In particular, there would be no effects due 
to field dependent mass terms in the kinetic part of the Lagrangian since they
arise in Skyrme--like models \cite{13} and lead to measure corrections
in the BFV--Hamiltonian path integral \cite{11,14} when integrating out the
field momenta. For these models, the equivalence of the Lagrangian BV and the
Hamiltonian BFV methods would have to be analysed again in detail.

Next we evaluate as a simple example of BV Lagrangian path integrals the
transition amplitude
\begin{eqnarray}
\langle\phi^f|\phi^i\rangle & = & \int^{\phi^f}_{\phi^i}
{\cal D}\{\eta(t,\x)\}{\cal D}\{a(t)\}{\cal D}\{c(t)\}{\cal D}\{\bar{c}(t)\}
{\cal D}\{b(t)\}\nonumber \\
& & \qquad \times\exp\left(\frac{i}{\hbar} 
S_{ext}^{\psi}[\eta,a,c,\bar{c},b;\eta^*=0,a^*=0,c^*=0,\bar{c}^*=0,b^*=0]\right)
\nonumber \\ 
& = & \int^{\phi^f}_{\phi^i}
{\cal D}\{\eta(t,\x)\}{\cal D}\{a(t)\}{\cal D}\{c(t)\}{\cal D}\{\bar{c}(t)\}
{\cal D}\{b(t)\}\nonumber \\
& & \qquad \times
\exp\bigg(\frac{i}{\hbar}\bigg\{S_0[\eta,a]  \nonumber \\
& & \qquad  \qquad {} - 
\int\bar{c}(t)\left[\int(\varphi'(\x-a(t)))^2d\x - 
\int\eta(t,\x)\varphi''(\x-a(t))d\x\right]c(t)dt \nonumber \\
& & \qquad\qquad {}- 
\int\left[\int\eta(t,\x)\varphi'(\x-a(t))d\x\right]b(t)dt\bigg\}\bigg).
\label{56}
\end{eqnarray}
The $b(t)$--integration yields 
\begin{equation}
\delta\left(\int\eta(t,\x)\varphi'(\x-a(t))d\x\right),
\label{delta}
\end{equation} and the 
$\bar{c}(t)c(t)$--integration results in the Faddeev--Popov like determinant
$\det_t(M_0+m_{\eta}(t))=M_0+{\cal O}(\eta)$ where
\begin{equation}
M_0=\int(\varphi'(\rho))^2d\rho
\label{57}
\end{equation}
is the classcal soliton mass and
\begin{equation}
m_{\eta}(t) = \int\eta(t,\x)\varphi''(\x-a(t))d\x = {\cal O}(\eta)
\label{58}
\end{equation}
is a quantum correction.

To handle the $\eta(t,\x)$ and $a(t)$ integrations, we expand $S_0[\eta,a]$
up to second order in powers of $\eta$:
\begin{eqnarray}
& & S_0[\eta,a] \nonumber \\& = & S_0[0,a] + \int 
\left[S_0[\eta,a]\frac{\dr}{\delta \eta(t,\x)}
\right]_{\eta=0}\eta(t,\x)dt d\x 
\nonumber \\
& & +{} \frac12\int\eta(t_1,\x_1)\left[
\frac{\dl}{\delta \eta(t_1,\x_1)}S_0[\eta,a]
\frac{\dr}{\delta \eta(t_2,\x_2)}
\right]_{\eta=0}\eta(t_2,\x_2)dt_1 d\x_1 dt_2 d\x_2
+ {\cal O}(\eta^3). \nonumber\\
\label{expansion}
\end{eqnarray}
The constant contribution is
\begin{equation}
S_0[0,a] = \tilde{S}_0[\varphi] + \int\frac12 M_0 \dot{a}^2(t)dt,
\label{59}
\end{equation}
the action of the classical soliton solution plus the action of the free
motion of a particle with mass $M_0$.

Due to the fact that $\varphi$ solves the static classical Euler--Lagrange 
equations for $\tilde{\LD}$, the terms linear in $\eta$ are
\begin{eqnarray}
& & \left[S_0[\eta,a]\frac{\dr}{\delta \eta(t,\x)}
\right]_{\eta=0}\eta(t,\x)dt d\x  =  \int \left[
\frac{\partial^2}{\partial t^2}\varphi(\x-a(t))\right]\eta(t,\x)dtd\x
\nonumber \\
& = & \int \ddot{a}(t) \left[\int \eta(t,\x)\varphi'(\x-a(t))d\x\right]dt
- \int \dot{a}^2(t)\left[\int \eta(t,\x)\varphi''(\x-a(t))d\x\right]dt.
\nonumber \\
\label{60}
\end{eqnarray}
The first term is cancelled by the $\delta$--function from the 
$b(t)$--integration (\ref{delta}), and the second yields a mass correction of
$2m_{\eta}(t)={\cal O}(\eta)$ to the free motion part of (\ref{59}). This
effect is another motivation for the choice of the gauge--fixing by eq. 
(\ref{52}): 
The gauge fermion is chosen such that only ``physically meaningful''
terms survive in the fluctuation expansion of the classical action. Another
possible gauge fixing would be to eliminate the complete linear contribution
in the $\eta$--expansion: The gauge fermion
\begin{equation}
\psi[\eta,a,\bar{c}] = 
\int \left[\eta(t,\x)
\frac{\partial^2}{\partial t^2}\varphi(\x-a(t))d\x\right]\bar{c}(t)dt
\label{61}
\end{equation}
could be a candidate for that idea.

What remains are the quadratic terms yielding harmonic quantum fluctuations in 
the usual way:
\begin{eqnarray}
& & \frac12\int\eta(t_1,\x_1)\left[
\frac{\dl}{\delta \eta(t_1,\x_1)}S_0[\eta,a]
\frac{\dr}{\delta \eta(t_2,\x_2)}
\right]_{\eta=0}\eta(t_2,\x_2)dt_1 d\x_1 dt_2 d\x_2
\nonumber \\
& = & \frac12\int\eta(t,\x)\left[-\frac{\partial^2}{\partial t^2}
+\frac{\partial^2}{\partial x^2} - V''(\varphi(\x-a(t)))\right]\eta(t,\x)dtd\x
\label{62a}
\end{eqnarray}
The substitution 
\begin{eqnarray}
\x & \mapsto & \rho=\x-a(t) \nonumber \\
t & \mapsto & t
\label{subst}
\end{eqnarray}
does not change the integral measure since its Jacobi determinant is $1$.
Under this change of variables, the second derivatives change to
\begin{equation}
\frac{\partial}{\partial t}\mapsto\frac{\partial}{\partial t}-\dot{a}(t)
\frac{\partial}{\partial \rho}, \qquad 
\frac{\partial}{\partial \x}\mapsto\frac{\partial}{\partial \rho}
\label{62b}
\end{equation}
and the harmonic fluctuation integral is
\begin{equation}
\frac12\int\eta(t,\rho+a(t))\left[-\frac{\partial^2}{\partial t^2} -\hat{\Omega}
\right]\eta(t,\rho+a(t))dtd\rho + 
\int m_{\eta^2}(t)\left(\dot{a}(t)\right)^2dt + S_0^{(int)}[\eta,a]
\label{62}
\end{equation}
where $\hat{\Omega}$ is the Gaussian fluctuation operator
\begin{equation}
\hat{\Omega}=-\frac{\partial^2}{\partial \rho^2} + V''(\varphi(\rho)).
\label{63}
\end{equation}
The change of variables also yields a further quantum correction to the soliton
mass (resulting in a modification of the center--of--mass action of the soliton)
by
\begin{equation}
m_{\eta^2}(t)=\int\left(\eta'(t,\rho+a(t))\right)^2d\rho = {\cal O}(\eta^2)
\label{62d}
\end{equation}
and $S_0^{(int)}[\eta,a]$ are coupling terms of second order in the 
fluctuations:
\begin{eqnarray}
& & \nonumber \\
S_0^{(int)}[\eta,a] & = & 
\frac12\int\eta(t,\rho+a(t))\left[\dot{a}(t)\frac{\partial}{\partial t}
\frac{\partial}{\partial \rho} +  \frac{\partial}{\partial t}\dot{a}(t)
\frac{\partial}{\partial \rho}
\right]\eta(t,\rho+a(t))dtd\rho\nonumber \\
& = & {\cal O}(a,\eta^2).
\label{62c}
\end{eqnarray}

It is easy to check that $\Psi_0(\rho)=\frac{1}{\sqrt{M_0}}\varphi'(\rho)$ is
a normalised zero mode of the Gaussian fluctuation operator, 
$\hat{\Omega}\Psi_0(\rho)=0$. Assuming
that we know a complete set of (generalised) eigenfunctions of $\hat{\Omega}$,
$\hat{\Omega}\Psi_n(\rho)=\omega_n^2\Psi_n(\rho)$, 
$\int\overline{\Psi_n(\rho)}\Psi_m(\rho)d\rho=\delta_{nm}$, 
we may expand the fluctuations in terms of $\Psi_n$:
\begin{equation}
\eta(t,\rho+a(t)) = \sum_n\alpha_n(t)\Psi_n(\rho).
\label{64}
\end{equation}
Eq.\ (\ref{64}) may be regarded as a linear substitution in the path integral 
(\ref{56}) with Jacobi determinant
\begin{equation}
\det\left(\frac{\partial \eta(t,\rho+a(t))}{\alpha_n(t)}\right)
 = J + {\cal O}(\eta,a)
\label{65}
\end{equation}
with constant $J$: Physically speaking, we changed the basis in the
space of the ``fluctuation coordinates'' from the ``continuous space index''
$\x$ to a quantum number index $n$: $a_{\x}(t)\mapsto \alpha_n(t)$, which
from the mathematical point of view are the Fourier series coefficients of
$\eta(t,\x)$ in terms of the basis of eigenfunctions.

Inserting (\ref{64}) into (\ref{62}), we obtain
\begin{equation}
\sum_{n=0}^{\infty}\int\left\{\frac12\dot{\alpha}^2_n - \frac12\omega_n^2
\alpha_n^2(t)\right\}dt
\label{66}
\end{equation}
as quadratic contribution in the $\eta$--expansion. The zero mode $\omega^2_0=0$
is eliminated by the $\delta$--function in the integrand which in the
new $\alpha_n(t)$--variables reads:
\begin{equation}
\delta\left(\int\eta(t,\x)\varphi'(\x-a(t))d\x\right) = \frac{1}{\sqrt{M_0}}
\delta(\alpha_0(t)). 
\label{67}
\end{equation}

Putting everything together, we have in the one--loop approximation
\begin{eqnarray}
\langle\phi^f|\phi^i\rangle & = & \int^{\phi^f}_{\phi^i}
\prod_{n=0}^{\infty}{\cal D}\{\alpha_n(t)\}{\cal D}\{a(t)\}
\left(J + {\cal O}(\eta,a)\right)\frac{1}{\sqrt{M_0}}\delta(\alpha_0(t))
\left(M_0+{\cal O}(\eta)\right) \nonumber \\
& & \qquad {} \times
\exp\bigg(\frac{i}{\hbar}\bigg\{
\tilde{S}_0[\varphi] + \int\frac12 \left(M_0+2m_{\eta}(t)
+2m_{\eta^2}(t)\right) \dot{a}^2(t)dt
\nonumber \\
& & \qquad\qquad\qquad {}
+\sum_{n=0}^{\infty}\int\left\{\frac12\dot{\alpha}^2_n - \frac12\omega_n^2
\alpha_n^2(t)\right\}dt + {\cal O}(\eta^3)+ {\cal O}(\eta^2,a)\bigg)\bigg\}
\nonumber \\ \label{68}
\end{eqnarray}

If we neglect quantum corrections to the soliton mass and to the Jacobian in
lowest order, the path integral factorises into three parts:
The action of the classical solution as a constant factor, 
$\exp\left(\frac{i}{\hbar}\tilde{S}_0[\varphi]\right)$, the free motion
of the center--of--mass of the soliton,
\begin{equation}
\int{\cal D}\{a(t)\}\exp\left(\frac{i}{\hbar}
\int\frac12 M_0 \dot{a}^2(t)dt\right)
\label{69}
\end{equation}
and the contribution of harmonic quantum fluctuations
\begin{equation}
\prod_{n\neq 0} \int{\cal D}\{\alpha(t)\}\exp\left(\frac{i}{\hbar}
\bar{S}_n[\alpha]\right),
\label{70}
\end{equation}
where
\begin{equation}
\bar{S}_n[\alpha] = \int\left\{\frac12\dot{\alpha}^2_n - \frac12\omega_n^2
\alpha_n^2(t)\right\}dt
\label{71}
\end{equation}
is the action of a harmonic oscillator with frequency $\omega_n$. The complete
integral is purely Lagrangian and avoids the mixed Lagrangian--Hamiltonian
notation which is often used in this context \cite{4,9}.

\section{The reduced $O(3)$ $\sigma$--model in $1+1$ dimensions,
collective rotation coordinates and their zero modes}
\label{section4}

In the previous sections, we investigated the BV quantisation of a model with 
one collective coordinate $a(t)$, describing a global translation symmetry
which is broken by the soliton solution $\varphi(x)$.

In general the collective coordinate method deals with theories given by
an action $S[\Phi]$ which is invariant under the action of a symmetry group
${\bf G}$ on the fields $\Phi^i(x)$. Static soliton solutions $\varphi^i(\x)$
of the Euler--Lagrange equations $S[\Phi]\frac{\dr}{\delta \Phi^i(x)}$ break
this symmetry group to a subgroup ${\bf H}\subset{\bf G}$, and there are zero 
modes associated with each element of ${\bf G}/{\bf H}$ \cite{9,21}. Collective
coordinates, in some sense, ``restore'' the broken symmetries \cite{22}
by introducing a new, {\em local} symmetry into the theory.

Translation invariance is the symmetry most commonly broken by soliton 
solutions: This is clear fron the fact that solitons must have finite
classical energy and thus are localised static objects in space. But besides
translation invariance which is a spacetime symmetry, internal symmetries
are important properties of a large class of field theories. The breakdown
of internal symmetry degrees of freedom in soliton solutions also leads
to important applications of the BV scheme in the context of collective 
coordinates.

We investigate such an ``internal collective coordinate'' in the context
of th reduced nonlinear
$O(3)$ $\sigma$--model in $1+1$ dimensions discussed by Mottola and Wipf
\cite{16}. It is given by the Lagrange density
\begin{equation}
\LD(\vec{\phi},\partial_{\mu}\vec{\phi}) = \lambda_1
\langle\partial_{\mu}\vec{\phi},\partial^{\mu}\vec{\phi}\rangle
-\lambda_0 V(\phi^{(3)}),\quad (\vec{\phi})^2 = 1
\label{72}
\end{equation}
with bosonic $\vec{\phi}=(\phi^{(1)},\phi^{(2)},\phi^{(3)})$, 
$\{x^{\mu}\}=\{t,\x\}$. The bracket $\langle,\rangle$ in the following 
always denotes the appropriate canonical scalar product in the internal 
symmetry space. The potential
$ V(\phi^{(3)})=1-\phi^{(3)}$ breaks the internal
$O(3)$ symmetry of the kinetic energy term in the Lagrangian to an $O(2)$
symmetry around the $\phi^{(3)}$ axis. We use this model not only because it 
has important applications to baryon and lepton number violation calculations,
but also because its internal symmetry group is as simple as possible,
depending on only one parameter.

In a way different from that of \cite{16}, we parametrise the fields
$\vp$ by ``spherical parameter fields'' $\vec{G}=(G^{(1)},G^{(2)})$,
\begin{equation}
\vp=\left(\begin{array}{c}\sin G^{(1)}(t,\x) \cos G^{(2)}(t,\x) \\
\sin G^{(1)}(t,\x) \sin G^{(2)}(t,\x) \\
\cos G^{(1)}(t,\x)\end{array}\right),
\label{73}
\end{equation}
to get rid of the constraint $(\vec{\phi})^2 = 1$. This parametrisation is more
appropriate to the introduction of collective coordinates in the reduced 
$O(3)$ $\sigma$--model which have been discussed neither in \cite{16} nor in the
further investigations of this and related models \cite{23}.

In terms of the parameter fields $\vec{G}$, the Lagrangian density can be
written
\begin{equation}
\LD(\vec{G},\partial_{\mu}\vec{G}) = \lambda_1\langle\partial_{\mu}\vec{G},
\hat{g}(\vec{G})\partial^{\mu}\vec{G}\rangle - \lambda_0 V(\cos G^{(1)})
\label{74}
\end{equation}
with a field dependent mass matrix
\begin{equation}
\hat{g}(\vec{G})= \left(\begin{array}{cc} 1 & 0 \\ 0 & \sin^2 G^{(1)} 
\end{array}
\right)
\label{75}
\end{equation}
which defines a Riemannian metric in the internal symmetry space of the
parameter fields. 
Eq.\ (\ref{74}) is a Lagrangian of the type discussed in \cite{9}.
It is invariant under spacetime transformations 
$x^{\mu}\mapsto x^{\mu}+ x^{\mu}_0$ and constant shifts of the second paramter 
field, $G^{(2)}\mapsto G^{(2)}+\gamma$, the latter corresponding to an $O(2)$
rotation of $\vec{\phi}$ around $\phi^{(3)}$ with angle $\gamma$.

It is well known that the ansatz $G^{(1)}_0(t,\x)=g(\x)$, 
$G^{(2)}_0(t,\x)=\gamma=\mbox{const}$ yields the Euler--Lagrange equation
\begin{equation}
2g''(\x) - \mu^2 \sin g(\x) = 0, \qquad \mu=
\pm\sqrt{\frac{\lambda_0}{\lambda_1}}
\label{76}
\end{equation}
with solution $g(\x)=4\arctan\exp\left(\frac{\mu}{\sqrt{2}}\x\right)$.
This corresponds to the sphaleron configuration
\begin{equation}
\vec{\varphi}(\x)=\left(\begin{array}{c}\sin g(\x) \cos \gamma \\
\sin g(\x) \sin \gamma \\
\cos g(\x)\end{array}\right) = 
\left(\begin{array}{c}
2\tanh\left(\frac{\mu}{\sqrt{2}}\x\right)
{\rm sech}\left(\frac{\mu}{\sqrt{2}}\x\right) \cos \gamma \\
2\tanh\left(\frac{\mu}{\sqrt{2}}\x\right)
{\rm sech}\left(\frac{\mu}{\sqrt{2}}\x\right) \sin \gamma \\
1-2{\rm sech}^2\left(\frac{\mu}{\sqrt{2}}\x\right)
\end{array}\right).
\label{77}
\end{equation}
We immediately see that $\vec{\varphi}(\x)$ breaks the space translation
invariance and the internal $O(2)$ symmetry of the Lagrangian (\ref{72}):
In terms of the parametrisation fields, $G^{(1)}_0(t,\x)=g(\x)$,
$G^{(2)}_0(t,\x)=\gamma$ and $G^{(1)}_0(t,\x)=g(\x-a^{(1)})$,
$G^{(2)}_0(t,\x)=\gamma+a^{(2)}$ are different solutions. (In the following,
we will use the solution with $\gamma=0$.) 

For each of these broken global symmetries, we now introduce one redundant
bosonic symmetry coordinate which we write as 
$\vec{A}(t)=(A^{(1)}(t),A^{(2)}(t))$. Since the Lagrangian density (\ref{74})
does not depend on $\vec{A}$ at all, we have two trivial Noether identities
\begin{equation}
\int\left(\tilde{S}_0[\vec{G},\vec{A}]\frac{\dr}{\delta G^{(i)}(t,\x)}\right)
R^{(G^{(i)})}d\x + 
\left(\tilde{S}_0[\vec{G},\vec{A}]\frac{\dr}{\delta A^{(i)}(t)}\right)
R^{(A^{(i)})} \equiv 0, \quad i=1,2
\label{78}
\end{equation}
with $R^{(G^{(i)})}=0$, $R^{(A^{(i)})}=1$. The corresponding extended action
in the field--antifield space $\{\tilde{Y}^A(x),\tilde{Y}^*_A(x)\} = 
\{\vG,\vA,\vC;\vGs,\vAs,\vCs\}$ (we added one fermionic ghost coordinate
$C^{(i)}(t)$ for each Noether identity) then reads
\begin{equation}
\tilde{S}_{ext}[\vec{G},\vec{A},\vec{C};\vec{G}^*,\vec{A}^*,\vec{C}^*]
=\tilde{S}_0[\vec{G},\vec{A}] + \int \langle\vAs,\vC\rangle dt.
\label{79}
\end{equation}

To introduce collective coordinates $\va=(a^{(1)}(t),a^{(2)}(t))$ and 
fluctuation fields
$\ve=(\eta^{(1)}(t,\x),\eta^{(2)}(t,\x))$  around the 
sphaleron solution $\vec{\varphi}(\x)$ given by 
$G^{(1)}_0(t,\x)=g(\x)$, $G^{(2)}_0(t,\x)=0$, we use a
canonical transformation
\begin{eqnarray}
& & \{\tilde{Y}^A(x),\tilde{Y}^*_A(x)\} = 
\{\vG,\vA,\vC;\vGs,\vAs,\vCs\} \nonumber \\
& \stackrel{F}{\mapsto} & \{Y^A(x),Y^*_A(x)\} = \{\ve,\va,\vc;\ves,\vas,\vcs\}
\label{80}
\end{eqnarray}
generated by
\begin{eqnarray}
F[\vec{G},\vec{A},\vec{C};\vec{\eta}^*,\vec{a}^*,\vec{c}^*] & = & 
\int\langle\vas,\vA\rangle dt + \int \langle\vcs,\vC\rangle dt \nonumber\\
& & {} + \int\eta^*_{(1)}(t,\x)\left[G^{(1)}(t,\x)-g\left(\x-A^{(1)}(t)\right)\right]dtd\x
\nonumber \\
& & {} + \int\eta^*_{(2)}(t,\x)\left[G^{(2)}(t,\x)-A^{(2)}(t)\right]dtd\x.
\label{81}
\end{eqnarray}
Applying (\ref{33}), we have $\vCs=\ves$, $\vCs=\vcs$, $\va=\vA$, $\vc=\vC$
and
\begin{eqnarray}
\eta^{(1)}(t,\x) & = & G^{(1)}(t,\x)-g\left(\x-A^{(1)}(t)\right) \label{82}\\
\eta^{(2)}(t,\x )& = & G^{(2)}(t,\x)-A^{(2)}(t) \label{83}\\
A^*_{(1)}(t) & = & a^*_{(1)}(t) + \int\eta^*_{(1)}(t,\x)g'\left(\x-A^{(1)}(t)\right)d\x
\label{84}\\
A^*_{(2)}(t) & = & a^*_{(2)}(t) - \int\eta^*_{(2)}(t,\x)d\x. \label{85}
\end{eqnarray}
Eq.\ (\ref{82}) and (\ref{83}) yield the transformation rules
\begin{eqnarray}
G^{(1)}(t,\x) & = & g\left(\x-a^{(1)}(t)\right) + \eta^{(1)}(t,\x) \label{86} \\
G^{(2)}(t,\x) & = & \quad a^{(2)}(t) \qquad \,\, + \eta^{(2)}(t,\x). \label{87}
\end{eqnarray}
Therefore, according to the definition (\ref{73}) of the parameter fields
$\vG$, we have to interpret $a^{(1)}(t)$ as 
{\em collective translation coordinate} and $\eta^{(1)}(t,\x)$ as fluctuations
in the ``direction of the sphaleron'', whereas (since $G^{(2)}(t,\x)$ is the 
angle of the $O(2)$ rotation around the $\phi^{(3)}$ axis in internal symmetry
space) $a^{(2)}(t)$ is a {\em collective rotation coordinate} and 
$\eta^{(2)}(t,\x)$ describes rotational fluctuations of the sphaleron around 
the fixed axis $\phi^{(3)}$.

Transforming the extended action (\ref{79}) via (\ref{80}) yields 
\begin{eqnarray} 
\Sev & = & S_0[\vec{\eta},\vec{a}]  \nonumber \\
& & {} + \int a^*_{(1)}(t)c^{(1)}(t)dt + 
\int\eta^*_{(1)}(t,\x)g'\left(\x-A^{(1)}(t)\right)c^{(1)}(t)dtd\x 
\nonumber \\
& & {} + \int a^*_{(2)}(t)c^{(2)}(t)dt + 
\int\eta^*_{(2)}(t,\x)(-1)c^{(2)}(t)dtd\x 
\label{88}
\end{eqnarray}
where
\begin{eqnarray}
& & 
S_0[\vec{\eta},\vec{a}] \nonumber \\
& = & \int\bigg\{\lambda_1\left[\left(-\dot{a}^{(1)}(t)
g'\left(\x-a^{(1)}(t)\right)+\dot{\eta}^{(1)}(t,\x)\right)^2 - \left(
g'\left(\x-a^{(1)}(t)\right)+\eta^{(1)\prime}(t,\x)\right)^2\right] \nonumber \\
& & {} + \lambda_1\left(\dot{a}^{(2)}(t)
+\dot{\eta}^{(2)}(t,\x)\right)^2 - \lambda_1\left(
\eta^{(2)\prime}(t,\x)\right)^2 
\sin^2\left( g\left(\x-a^{(1)}(t)\right) + \eta^{(1)}(t,\x)\right) \nonumber \\
& & {} -\lambda_0 V\left( g\left(\x-a^{(1)}(t)\right) + \eta^{(1)}(t,\x)\right)
\label{89}
\end{eqnarray}
is the new classical action in terms of collective and fluctuation degrees 
of freedom.

Comparing (\ref{88}) with the general ansatz for the extended action 
(\ref{ansatz}), we see that only terms linear in the antifields occur which 
means that the gauge algebra is again abelian. The BRST transformations can
be found by comparing (\ref{28}) and (\ref{88}), yielding
\begin{eqnarray}
\s \eta^{(1)}(t,\x) & = & g'\left(\x-a^{(1)}(t)\right)c^{(1)}(t) = R^{(\eta^{(1)})}
c^{(1)}(t) \nonumber \\
\s a^{(1)}(t) & = & c^{(1)}(t) = R^{(a^{(1)})}c^{(1)}(t) \nonumber \\
\s c^{(1)}(t) & = & 0 \label{90} \\[2mm]
\s \eta^{(2)}(t,\x) & = & -c^{(2)}(t) = R^{(\eta^{(2)})}
c^{(1)}(t) \nonumber \\
\s a^{(2)}(t) & = & c^{(2)}(t) = R^{(a^{(2)})}c^{(2)}(t) \nonumber \\
\s c^{(2)}(t) & = & 0 \label{91}
\end{eqnarray}
with Noether generators $R^{(\eta^{(1)})}= g'\left(\x-a^{(1)}(t)\right)$, 
$ R^{(a^{(1)})}=1$ which are typically associated with collective
translation coordinates (cf.\ the Noether generators (\ref{14}), (\ref{15})
of the simple scalar field theory model in section \ref{section1}), and
$R^{(\eta^{(2)})}=-1$, $R^{(a^{(2)})}=1$: These are typical Noether generators
associated with collective rotation degrees of freedom, they are related to the
generator of $O(2)$ rotations, $\left(\begin{array}{cc} 0 & -1 \\ 1 & 0
\end{array}\right)$.

It is easy to check that the Noether identities
\begin{equation} 
\int \left(S_0[\vec{\eta},\vec{a}]\frac{\dr}{\delta \eta^{(i)}(t,\x)}\right)
R^{(\eta^{(i)})}dx + 
\left(S_0[\vec{\eta},\vec{a}]\frac{\dr}{\delta  a^{(i)}(t)}\right)
R^{(a^{(i)})} \equiv 0
\label{92}
\end{equation}
hold for $i=1,2$: The $\x$--integration shows that the space position coordinate
is again a kind of ``continuous index'', so that the fluctuation fields can be
interpreted as ``fluctuation coordinates'' $\vec{a}_{\x}(t)=\vec{\eta}(t,\x)$.

To quantize the model by path integrals, we have to choose a convenient gauge
fermion $\psi$ and fix the gauge by performing the canonical transformation
$S_{ext}\stackrel{\psi}{\mapsto}S_{ext}^{\psi}$. Due to the fact that we now 
have two Noether identities (\ref{92}), we have to add two trivial systems
$\{\bar{c}^{(i)}(t),b^{(i)}(t)\}$, $i=1,2$ to construct the gauge fermion.

Without going into details concerning the evaluation of path integrals 
in the nonlinear reduced $O(3)$ $\sigma$ model, we finally consider the
zero modes and the
choice of gauge fixing in this model. From (\ref{90}) and (\ref{91}), we see 
that the translational and rotational gauge symmetries are independent of 
each other, and may therefore be gauge fixed independently. For the collective
translation degree of freedom, we know a suitable gauge fixing principle which
states that fluctuations parallel to the zero mode have to be excluded from
the path integration. Denoting the parallel and rotational zero modes
by $\Psi^{(1)}_0(\x)$ and 
$\Psi^{(2)}_0(\x)$ respectively, the corresponding gauge fermion reads
\begin{equation}
\psi[\vec{\eta},\vec{a},\vec{\bar{c}}]  =  
\sum_{i=1,2}
\int\bar{c}^{(1)}(t)\left\{\int\eta^{(i)}(t,\x)\Psi^{(i)}_0(\x-a^{(i)}(t))d\x
\right\}dt 
\label{93}
\end{equation}

From Section \ref{section3}, we may conclude that the translational zero mode
is $\Psi^{(1)}_0\propto g'$ and thus
$\Psi^{(1)}_0\propto R^{(\eta^{(1)})}$. Therefore, we may guess that the 
rotational zero mode is proportional to the corresponding Noether generator
$R^{(\eta^{(2)})}=-1$ which means that this zero mode is constant,
$\Psi^{(2)}_0\propto -1$ and thus seems to be not normalizable.

We can see in which sense this is true by investigating the expansion of the
classical action $S_0[\vec{\eta},\vec{a}]$ (\ref{89}) in terms of the 
fluctuations up to second order (which yields, after the $\vec{b}(t)$ and 
$\vec{c}(t)\vec{\bar{c}}(t)$ integrations which we skip here, the one loop
approximation in the path integral). This expansion may be written as
\begin{eqnarray}
S_0[\vec{\eta},\vec{a}] & = & \tilde{S}_0[\vec{\varphi}] +
S_0^{(CM1)}[\eta^{(1)},a^{(1)}] + S_0^{(CM2)}[\eta^{(1)},a^{(2)}]
\nonumber \\
& & \qquad \quad {} + S_0^{(int1)}[\eta^{(1)},a^{(1)}] + 
S_0^{(int2)}[\vec{\eta},\vec{a}] \nonumber \\
& & \qquad \quad {} + S_0^{(fluc1)}[\eta^{(1)},a^{(1)}] + 
\check{S}_0^{(fluc2)}[\eta^{(2)},a^{(1)}].
\label{94}
\end{eqnarray}

Here $\tilde{S}_0[\vec{\varphi}]=\int\left[-\lambda_1(g'(\rho))^2-\lambda_0
V(\cos g(\rho))\right] d\rho dt$ is the classical action of the sphaleron,
and
\begin{equation}
S_0^{(CM1)}[\eta^{(1)},a^{(1)}] = \lambda_1\int\left(\dot{a}^{(1)}\right)^2
[M-2m_{\eta^{(1)}}(t)]dt
\label{96}
\end{equation}
with $M=\int(g'(\rho))^2d\rho$ and 
\begin{equation}
m_{\eta^{(1)}}(t)=\int\eta^{(1)}(t,\x)g''\left(\x-a^{(1)}(t)\right)d\x = 
{\cal O}(\eta^{(1)})
\label{98}
\end{equation}
is the action of the free center--of--mass translation of the sphaleron,
whereas
\begin{equation}
S_0^{(CM2)}[\eta^{(1)},a^{(2)}] = \lambda_1\int\left(\dot{a}^{(2)}\right)^2
[I-2i_{\eta^{(1)}}(t)]dt
\label{99}
\end{equation}
with $I=\int\sin^2g(\rho)d\rho$ and
\begin{eqnarray}
i_{\eta^{(1)}}(t) & = & \int\left[
2\eta^{(1)}(t,\x)\cos g\left(\x-a^{(1)}(t)\right)
\sin g\left(\x-a^{(1)}(t)\right) \right.\nonumber \\
& & \qquad {} +\left. \left(2\eta^{(1)}(t,\x)\right)^2
\left(\cos^2 g\left(\x-a^{(1)}(t)\right) - \sin^2 g\left(\x-a^{(1)}(t)\right)
\right)\right] d\x
\nonumber \\
& = & {\cal O}(\eta^{(1)})
\label{101}
\end{eqnarray}
may be interpreted as the action of the rotation of the sphaleron around the
$\phi^3$ axis.

Next, we have mixed or interaction terms
\begin{eqnarray}
& & S_0^{(int1)}[\eta^{(1)},a{(1)}] = 
2\lambda_1\int\ddot{a}^{(1)}(t)
\left[\int\eta^{(1)}(t,\x) g'\left(\x-a^{(1)}(t)\right)d\x\right]dt \label{102}\\
& & S_0^{(int2)}[\vec{\eta},\vec{a}] = 2\lambda_1\int\!\!\!\int
\dot{a}^{(2)}(t)\dot{\eta}^{(2)}(t,\x)\sin^2 g\left(\x-a^{(1)}(t)\right)dtd\x \nonumber \\
& & \qquad\qquad\qquad {} + 4\lambda_1\int\!\!\!\int
\dot{a}^{(2)}(t)\eta^{(1)}(t,\x)\dot{\eta}^{(2)}(t,\x) \nonumber \\
& & \qquad\qquad\qquad\qquad\qquad {} \times
\cos g\left(\x-a^{(1)}(t)\right)\sin g\left(\x-a^{(1)}(t)\right)dtd\x 
\label{103}
\end{eqnarray}

The term (\ref{102}) equals the first integral in (\ref{60}) which was cancelled
by the gauge fixing condition. Since 
$\Psi^{(1)}_0\left(\x-a^{(1)}(t)\right)
\propto g'\left(\x-a^{(1)}(t)\right)$, a gauge fermion of the
type (\ref{93}) removes the interaction integral (\ref{102}) from the
expansion. The second interaction term (\ref{103}) is due to the collective
rotation coordinate and contains quite complicated couplings between all
fluctuations to all collective coordinates. One can try to choose the gauge 
fixing condition 
such that (\ref{103}) becomes as simple as possible, but it seems
obvious that it is impossible to cancel (\ref{103}) by gauge fixing.

Finally, we have the quadratic fluctuation parts of the classical action: The 
translational fluctuations $\eta^{(1)}$ yield, substituting
$\rho=\x-a^{(1)}(t)$ (which leads to a further interaction terms as discussed in
Section \ref{section3}),
\begin{eqnarray}
S_0^{(fluc1)}[\eta^{(1)},a^{(1)}] & = & 
\lambda_1\int\eta^{(1)}\left(t,\rho+a^{(1)}(t)\right)
\left[-\frac{\partial^2}{\partial t^2} - \hat{\Omega}^{(1)}\right]
\eta^{(1)}\left(t,\rho+a^{(1)}(t)\right)d\rho dt \nonumber \\
& & {} + S_0^{(int3)}[\eta^{(1)},a^{(1)}]
\label{104}
\end{eqnarray}

The interaction term is the analogue of (\ref{62d},\ref{62c}):
\begin{eqnarray}
& &
S_0^{(int3)}[\eta^{(1)},a^{(1)}] \nonumber \\
& = & \lambda_1\int\eta^{(1)}\left(t,\rho+a^{(1)}(t)\right)
\left[\dot{a}^{(1)}(t)\frac{\partial}{\partial t}
\frac{\partial}{\partial \rho} +
\frac{\partial}{\partial t}\dot{a}^{(1)}(t)
\frac{\partial}{\partial \rho}
\right]
\eta^{(1)}\left(t,\rho+a^{(1)}(t)\right)d\rho dt \nonumber \\
& & {} + \lambda_1
\int\left(\int\left(\eta^{(1)\prime}(t,\rho+a(t))\right)^2d\rho\right)
\left(\dot{a}^{(1)}(t)\right)^2 dt \nonumber \\
& = & {\cal O}\left((\eta^{(1)})^2,a^{(1)}\right)
\end{eqnarray}

The Gaussian fluctuation operator is
\begin{equation}
\hat{\Omega}^{(1)}=-\frac{\partial^2}{\partial\rho^2} +
\frac12\mu\left[\sin^2 g(\rho) V''(\cos g(\rho)) - 
\cos g(\rho)V'(\cos g(\rho))\right]
\label{105}
\end{equation}
Inserting the potential $V(\phi^{(3)})=1-\phi^{(3)}$ and the
appropriate solution $g$, this yields (with $y=\frac{\rho}{\sqrt{2}}$)
\begin{equation}
\hat{\Omega}^{(1)} = \frac{\mu^2}{2}\left[\frac{\partial^2}{\partial y^2}
+(1-2{\rm sech}^2 y)\right]
\label{106}
\end{equation}
which is the fluctuation operator ``parallel to the sphaleron'' discussed
by Mottola and Wipf \cite{16}.

From the rotational fluctuation, we have in the one loop approximation
\begin{equation}
\check{S}_0^{(fluc2)}[\eta^{(2)},a^{(1)}] = \lambda_1 \int\eta^{(2)}(t,\x)
\left[-\frac{\partial}{\partial t}\sin^2 g\left(\x-a^{(1)}(t)\right)
\frac{\partial}{\partial t} - \check{\Omega}^{(2)}\right]\eta^{(2)}(t,\x)d\x dt
\label{107}
\end{equation}
with
\begin{equation}
\check{\Omega}^{(2)} = -\frac{\partial}{\partial \x}\sin^2 g\left(\x-a^{(1)}(t)\right)
\frac{\partial}{\partial \x}
\label{108}
\end{equation}
This Gaussian fluctuation operator has a constant zero mode, 
$\check{\Omega}^{(2)} 1=0$, and the corresponding gauge fixing condition for
the $\eta^{(2)}$ fluctuations would be $\int\eta^{(2)}(t,\x)\cdot 1 d\x =0$.

Although this zero mode is a constant function, one can not conclude that
it is not normalizable (a fact from which would follow that the zero mode is
part of the continuous spectrum): The eigenvalue problem we have to solve
is not simply $\check{\Omega}^{(2)}\Psi=\omega^2\Psi$, but we have to include
a measure function on the right hand side \cite{24}, 
because the time derivative
in (\ref{107}) does not have the simple form $-\frac{\partial^2}{\partial t^2}$
which is needed to decompose the fluctuations in harmonic oscillations as shown
in Section \ref{section3}. The measure funtion has to be the (time independent)
prefactor of the pure second order time derivative in (\ref{107}). One can check
that in terms of the new integration variable $\rho=\x-a^{(1)}(t)$, the
integral (\ref{107}) becomes
\begin{eqnarray}
& &
\check{S}_0^{(fluc2)}[\eta^{(2)},a^{(1)}] \nonumber \\
& = & \lambda_1
\int\eta^{(2)}\left(t,\rho+a^{(1)}(t)\right)\nonumber \\
& & \qquad\qquad\times
\left[-\sin^2 g(\rho)\frac{\partial^2}{\partial t^2}
+\frac{\partial}{\partial \rho}\sin^2 g(\rho)\frac{\partial}{\partial \rho}
\right]\eta^{(2)}\left(t,\rho+a^{(1)}(t)\right) dt d\rho \nonumber \\
& & {} + S_0^{(int4)}[\eta^{(2)},a^{(1)}]
\label{108a}
\end{eqnarray}
with
\begin{eqnarray}
& & 
S_0^{(int4)}[\eta^{(2)},a^{(1)}] \nonumber \\
& = & \lambda_1
\int\eta^{(2)}\left(t,\rho+a^{(1)}(t)\right)
\left[\dot{a}^{(1)}(t)\frac{\partial}{\partial \rho}\sin^2 g(\rho)
\frac{\partial}{\partial t} \right. \nonumber \\
& & \qquad\qquad\qquad\qquad\qquad {} + \left.
\sin^2 g(\rho)\frac{\partial}{\partial t}\dot{a}^{(1)}(t)
\frac{\partial}{\partial \rho}
\right]\eta^{(2)}\left(t,\rho+a^{(1)}(t)\right)dt d\rho \nonumber \\
& & {} - \lambda_1\int\left\{\int\eta^{(2)}\left(t,\rho+a^{(1)}(t)\right)
\sin^2 g(\rho)\eta^{(2)\prime\prime}\left(t,\rho+a^{(1)}(t)\right)d\rho\right\}
\left(\dot{a}^{(1)}(t)\right)^2 dt \nonumber \\
& = & {\cal O}\left((\eta^{(2)})^2,a^{(1)}\right),
\end{eqnarray}
so the measure function to
be inserted into the eigenvalue equation is
$r(\rho)=\sin^2 g(\rho)=4\tanh^2\left(\frac{\mu}{\sqrt{2}}\x\right)
{\rm sech}^2\left(\frac{\mu}{\sqrt{2}}\x\right)$. 
The eigenvalue problem one has to solve is thus
\begin{equation}
-\frac{\partial}{\partial \rho}r(\rho)\frac{\partial}{\partial \rho}\Psi(\rho)
= \omega^2r(\rho)\Psi(\rho)
\end{equation}
which is a Sturm--Liouville problem. The eigenfunctions are orthogonal
with respect to the inner product 
$(\Psi_a,\Psi_b)_r=
\int r(\rho)
\overline{\Psi_a(\rho)}\Psi_b(\rho)d\rho$, and in terms of this inner
product, the constant zero mode of $\check{\Omega}^{(2)}$ is normalizable: 
$(1,1)_r<\infty$.

One can simplify the complicated structure 
of the rotational fluctuation operator
by changing the fluctuations to
\begin{equation}
\xi\left(t,\rho+a^{(1)}(t)\right)=
\sin g(\rho)\cdot\eta^{(2)}\left(t,\rho+a^{(1)}(t)\right).
\end{equation}
Inserting this substitution into (\ref{108a}), we obtain
\begin{eqnarray}
& & \lambda_1\int\eta^{(2)}\left(t,\rho+a^{(1)}(t)\right)\nonumber \\
& & \qquad\qquad\times
\left[-\sin^2 g(\rho)\frac{\partial^2}{\partial t^2}
+\frac{\partial}{\partial \rho}\sin^2 g(\rho)\frac{\partial}{\partial \rho}
\right]\eta^{(2)}\left(t,\rho+a^{(1)}(t)\right) dt d\rho \nonumber \\
& = & 
\lambda_1\int\xi\left(t,\rho+a^{(1)}(t)\right)
\left[-\frac{\partial^2}{\partial t^2} + \hat{\Omega}^{(2)}\right]
\xi\left(t,\rho+a^{(1)}(t)\right)dtd\rho \nonumber \\
& = & S_0^{(fluc2)}[\xi,a^{(1)}]
\label{115}
\end{eqnarray}
with 
\begin{equation}
\hat{\Omega}^{(2)}=-\frac{\partial^2}{\partial \rho^2}
+\left(g''(\rho)\cot g(\rho) - (g(\rho))^2\right)
\label{116}
\end{equation}
which yields, inserting the sphaleron solution $g(\rho)$ and setting
$y=\frac{\mu}{\sqrt{2}}\rho$:
\begin{equation}
\hat{\Omega}^{(2)} = \frac{\mu^2}{2}\left[ - \frac{\partial^2}{\partial y^2}
+(1-6{\rm sech}^2y)\right]
\label{117}
\end{equation}
This is the the second P\"oschl-Teller operator discussed in \cite{16,23}.
From $\check{\Omega}^{(2)} 1 =0$ we have the zero mode
$\hat{\Omega}^{(2)}\Psi_0^{(2)}=0$ with 
\begin{equation}
\Psi_0^{(2)}(y)=\sin g(y)\cdot 1 = 
2\tanh\left(\frac{\mu}{\sqrt{2}}\x\right)
{\rm sech}^2\left(\frac{\mu}{\sqrt{2}}\x\right)
\label{118}
\end{equation}
which is square integrable in the usual sense: There is no more
prefactor to the second time derivative in (\ref{115}). We also see that
the zero mode  
eigenfunction (\ref{118}) has one node, so there must be a ground state 
eigenfunction with no node which has a negative eigenvalue.

\section{Conclusions}

We have shown how the breakdown of global symmetries of a given Lagrangian
in its soliton or sphaleron solutions can be cured by the introduction of
collective coordinates. The Lagrangian field--antifield formalism of
Batalin and Vilkovisky proved to be an appropriate tool to perform the 
well-known main steps of the collective coordinate method, such as the 
introduction of collective and fluctuation degrees of freedom, the 
investigation of the resulting gauge algebra structure related to the
Noether identities of the theory, the BRST treatment of the gauge symmetry,
finally the choice of gauge fixing and the evaluation of path integrals.
We also used the simple gauge structure of soliton models to explore some
pecularities of the BV scheme in concrete detail, in particular canonical
transformations were used to introduce collective coordinates and to fix the 
gauge.

Although the BV scheme clarifies the collective coordinate method from a 
structural point of view, the explicit evaluation of path integrals remains
a complicated task: The special mapping from the original fields to
collective coordinates and fluctuation fields (\ref{2}) was used throughout
this article since it may be derived from a canonical transformation, and yields
complicated interaction terms when we expand the classical action up to
harmonic fluctuations. Some of these vanish due to an appropriate choice of
the gauge fixing condition, others may be interpreted as quantum correction to 
the soliton mass.

We did not go beyond the one loop approximation in the path integral. In fact,
it is possible to write down the terms contributing to higher loop
expansions, but it will be more difficult 
to evaluate the corresponding path integrals.

\section*{Acknowledgements}

This research has been carried out in the framework of the European Human 
Capital and Mobility Network on ``Constrained Dynamical Systems''. The
author is indebted to F. de Jonghe, J.--Q. Liang, H.J.W. M\"uller--Kirsten
and J.--G. Zhou for helpfull discussions.

\end{document}